\documentclass[pre,10pt,twocolumn,preprint]{revtex4} %[aps,preprint]{revtex4}
\usepackage[dvips]{graphicx}
\usepackage{latexsym}
\usepackage{natbib}
\usepackage[centerlast]{caption}
\usepackage{subfig}

\begin{document}
\title{Robustness of a Tree-like Network of Interdependent Networks (28 August)}

\author{Jianxi Gao,$^{1,2}$ S. V. Buldyrev,$^3$ S. Havlin,$^4$ and
H. E. Stanley$^2$}

\affiliation{$^1$Department of Automation, Shanghai Jiao Tong
University, 800 Dongchuan Road, Shanghai, 200240, PR China\\
$^2$Center for Polymer Studies and Department of
Physics, Boston University, Boston, MA 02215 USA\\
$^3$Department of Physics,~Yeshiva University, New York, NY 10033 USA\\
$^4$Department of Physics, Bar-Ilan University, 52900 Ramat-Gan,
Israel}

\date{\bf \today --- gbhs28AugustPRL.tex}

\begin{abstract}
In reality, many real-world networks interact with and depend on
other networks. We develop an analytical framework for studying
interacting networks and present an exact percolation law for a
network of $n$ interdependent networks (NON). We present a general
framework to study the dynamics of the cascading failures process at
each step caused by an initial failure occurring in the NON system.
We study and compare both $n$ coupled Erd\H{o}s-R\'{e}nyi (ER)
graphs and $n$ coupled random regular (RR) graphs. We found recently
[Gao et. al. arXive:1010.5829] that for an NON composed of $n$ ER
networks each of average degree $k$, the giant component,
$P_{\infty}$, is given by $P_{\infty}=p[1-\exp(-kP_{\infty})]^n$
where $1-p$ is the initial fraction of removed nodes. Our general
result coincides for $n=1$ with the known Erd\H{o}s-R\'{e}nyi
second-order phase transition at a threshold, $p=p_c$, for a single
network. For $n=2$ the general result for $P_{\infty}$ corresponds
to the $n=2$ result [Buldyrev et. al., Nature , 464, (2010)]. Here
we show for an NON composed of $n$ coupled RR networks each of
degree $k$, that the giant components is given by $ P_{\infty} = p
\big \{1-\{p^{1/n}P_{\infty}^{(n-1)/n} [(1-(P_{\infty}/p )^{1/n}
)^{(k-1)/k}-1 ]+1\}^k \big \}^n$. Similar to the ER NON, for $n=1$
the percolation transition at $p_c$, is of second order while for
any $n>1$ it is of first order. The first order percolation
transition in both ER and RR (for $n>1$) is accompanied by cascading
failures between the networks due to their interdependencies.
However, we find that the robustness of $n$ coupled RR networks of
degree $k$ is dramatically higher compared to the $n$ coupled ER
networks of average degree $k$. While for ER NON there exists a
critical minimum average degree $k= k_{\min}$, that increases with
$n$, below which the system collapses, there is no such analogous
$k_{\min}$ for RR NON system. For any $k>2$, the RR NON is stable,
i.e., $p_c<1$. This is due to the critical role played by singly
connected nodes which exist in ER NON and enhance the cascading
failures but do not exist in the RR NON system.
\end{abstract}
\maketitle
\section{Introduction}

In our modern world, infrastructures, which affect all areas of
daily life, are usually interdependent. Examples include, electric
power, natural gas and petroleum production and distribution,
telecommunications, transportation, water supply, banking and
finance, emergency and government services, agriculture, and other
fundamental systems and services that are critical to the security,
economic prosperity, and social systems, shown in Fig.1. Although
urban societies rely on each of the individual infrastructures,
recent disasters ranging from hurricanes to large-scale power
blackout and terrorist attacks have shown that significant dangerous
vulnerability is due to the many interdependencies across different
infrastructures~\cite{Chang2009,Rinaldi2001,Alessandro2010,John2008,Rosato2008}.
Infrastructures are frequently connected at multiple points through
a wide variety of mechanisms, such that a bidirectional relationship
exists between the states of any given pair of networks, as shown in
Fig.~1 and 2~\cite{Rinaldi2001}. For example, in California,
electric power disruptions in early 2001 affected oil and natural
gas production, refinery operations, pipeline transport of gasoline
and jet fuel within California and its neighboring states, and the
movement of water from northern to central and southern regions of
the state for crop irrigation. Another dramatic real-world example
of a cascade of failures is the electrical blackout that affected
much of Italy on 28 September 2003: the shutdown of power stations
directly led to the failure of nodes in the Supervisory Control and
Data Acquisition (SCADA) communication network, which in turn caused
further breakdown of power stations~\cite{Rosato2008,Sergey2010}.
Identifying, understanding, and analyzing such interdependencies are
significant challenges. These challenges are greatly magnified by
the breadth and complexity of our modern critical national
interdependent infrastructures~\cite{John2008}.

In recent years we observed important advances in the field of
complex networks
~\cite{Strogatz1998,bara2000,Callaway2000,Albert2002,Cohen2000,Newman2003,
Dorogovtsev2003,song2005,Pastor2006,Caldarelli2007,Barrat2008,Shlomo2010,Neman2010}.
The internet, airline routes, and electric power grids are all
examples of networks whose function relies crucially on the
connectivity between the network components. An important property
of such systems is their robustness to node failures. Almost all
research has been concentrated on the case of a single or isolated
network which does not interact with or depend on other networks.
Recently, based on the motivation that modern infrastructures are
becoming significantly more dependent on each other, a system of two
coupled interdependent networks has been studied~\cite{Sergey2010,
parshani2010,Wei2011}. A fundamental property of interdependent
networks is that when nodes in one network fail, they may lead to
the failure of dependent nodes in other networks which may cause
further damage in the first network and so on, leading to a global
cascade of failures. Buldyrev et al.~\cite{Sergey2010} developed a
framework for analyzing the robustness of two interacting networks
subject to such cascading failures. They found that interdependent
networks behave very different from single networks and become
significantly more vulnerable compared to their noninteracting
counterparts.

In many realistic examples, more than two networks depend on each
other. For example, diverse infrastructures, such as water and food
supply, communications, fuel, financial transactions, and power
stations are coupled
together~\cite{Peerenboom2001,Rinaldi2001,Rosato2008,Alessandro2010}.
Understanding the vulnerability due to such interdependencies is a
major challenges for designing resilient infrastructures.

\begin{figure}[h!]
\centering \includegraphics[width=0.48\textwidth]{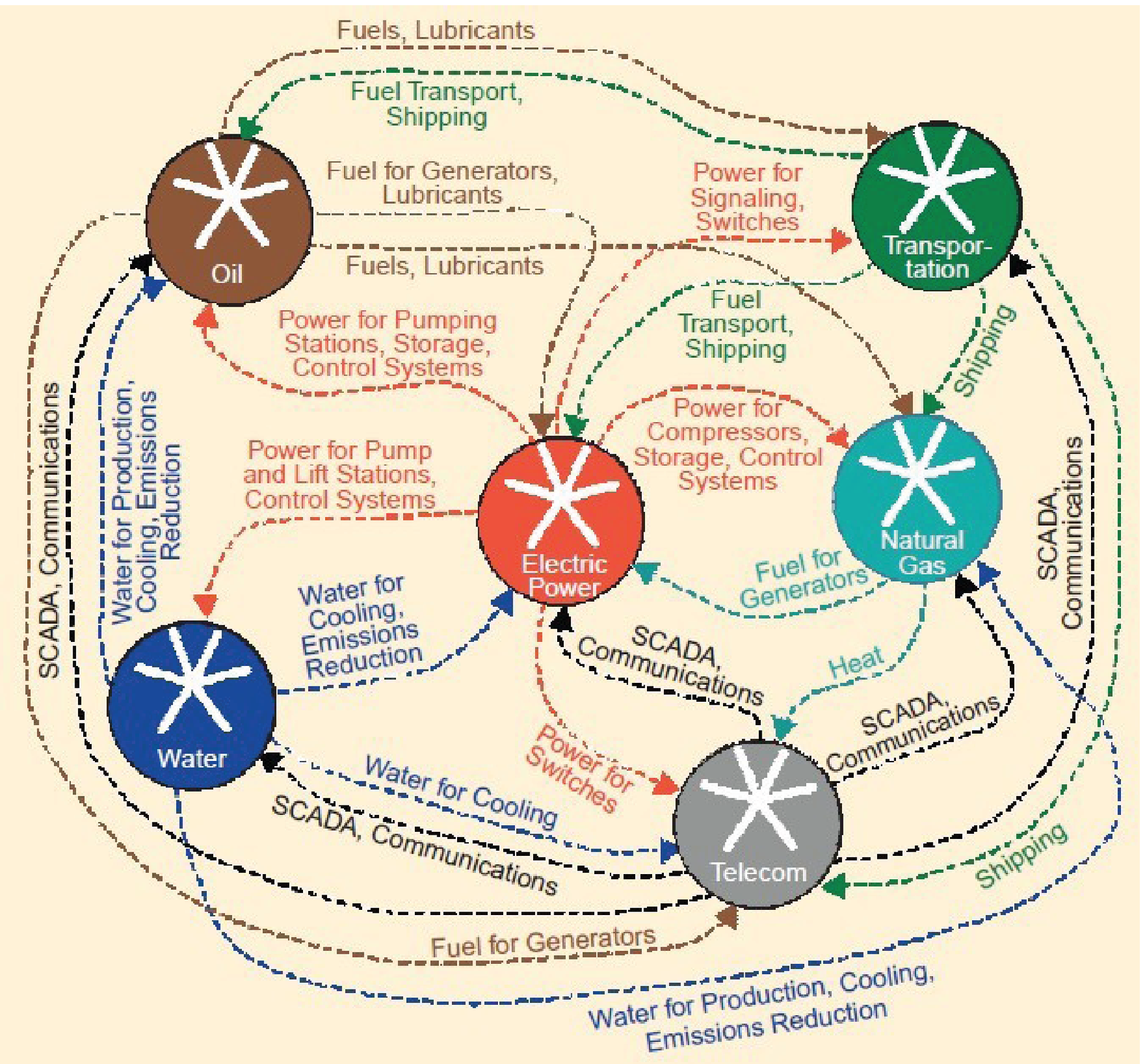}
\caption{Illustration of the interdependent relationship among
different infrastructures~\cite{Rinaldi2001}. These complex
relationships are characterized by multiple connections between
infrastructures, feedback and feedforward paths, and intricate,
branching topologies. The connections create an intricate web that,
depending on the characteristics of its linkages, can transmit
shocks throughout broad swaths of an economy and across multiple
infrastructures. It is clearly impossible to adequately analyze or
understand the behavior of a given infrastructure in isolation from
the environment or other infrastructures. Rather, one must consider
multiple interconnected infrastructures and their interdependencies.
For example, the reliable operation of modern infrastructures
depends on computerized control systems, from SCADA systems that
control electric power grids to computerized systems that manage the
flow of railcars and goods in the rail industry. In these cases, the
infrastructures require information transmitted and delivered by the
communication
infrastructure~\protect\cite{Rinaldi2001}.}\label{fig1}
\end{figure}

We study here a model system ~\cite{gao2010}, comprising a network
of $n$ coupled networks, where each network consists of $N$ nodes
(See Fig.~2). The $N$ nodes in each network are connected to nodes
in neighboring networks by bidirectional dependency links, thereby
establishing a one-to-one correspondence. We apply a mathematical
framework \cite{gao2010} to study the robustness of tree-like
``network of networks'' (NON) by studying the dynamically process of
the cascading failures. We find an exact analytical law for
percolation of a NON system composed of $n$ coupled randomly
connected networks. Our result generalizes the known
Erd\H{o}s-R\'{e}nyi (ER)~\cite{ER1959,ER1960,Bollob1985} result as
well as the random regular (RR) result for the giant component of a
single network, and shows that while for $n=1$ the percolation
transition is a second order, for $n>1$ cascading failures occur and
the transition becomes a first order transition. Our results for $n$
interdependent networks show that the classical percolation theory
extensively studied in physics and mathematics is in fact a limited
case of the rich, general, and very different percolation law which
exists in realistic interacting networks.

Additionally, we find:

(i) for any loopless topology of NON, the critical percolation
threshold and the giant component depend only on the {\it number\/}
of networks involved and their degree distributions but {\it not on
the inter-linked topology\/} (Fig.~2),

(ii) the robustness of NON {\it significantly decreases\/} with $n$,
and

(iii) for a network of $n$ ER networks all with the same average
degree $k$, there exists a minimum degree $k_{\min}(n)$ increasing
with $n$, below which $p_c =1$, i.e., for $k<k_{\min}$ the NON will
collapse once any finite number of nodes fail. The analytical
expression for $k_{\min}(n)$ generalizes the known result
$k_{\min}(1)=1$ for ER below which the network collapses. In sharp
contrast a NON composed of RR networks is significantly more robust.
In the RR NON case there is no $k_{\min}$ below which the NON
collapses. This is due to the multiple links of each node in the RR
system compared to the existence of singely connected nodes in the
ER case. We also discuss the critical effect of singly connected
nodes on the vulnerability of the NON ER structure.

\begin{figure}[h!]
\centering
\includegraphics[width=0.48\textwidth]{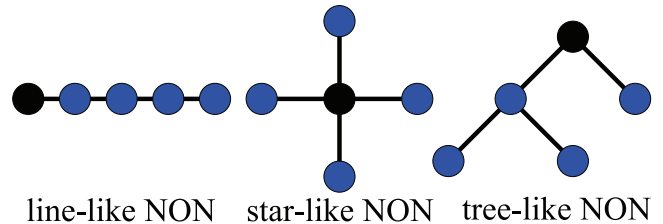}
\caption{(color online) Three types of loopless NONs composed of
five coupled networks all have same percolation threshold and same
giant percolation component.}\label{fig2}
\end{figure}

\section{The dynamic process of cascading failures}

To model an interdependent NON we consider for simplicity and
without loss of generality, $n$ networks each having $N$ nodes. We
study the percolation of $n$ networks connected in a loopless
structure, the structure of the NON can be, e.g., a line, a star or
a tree as shown in Fig.~2. Each node in Fig.~2 represents a network,
and each link between two networks $i$ and $j$ denotes the existence
of a one-to-one dependencies between the nodes of the linked
networks. The functioning of one node in network $i$ depends on the
functioning of one and only one node in network $j$ ($i,j \in
\{1,2,...,n\}, i\neq j$), and vice versa (bidirectional links). We
assume that within network $i$, the nodes are randomly connected by
$A_i$-links with degree distribution $P(k_i)$, where $k_i$ is the
average degree of network $i$.

The root of the NON is the network from which fraction $1-p$ of
nodes are removed due to random failure. Before showing the dynamic
of the cascading failures, we present the following three
definitions. (i) We define the distance matrix $D_{ij}$ as the
distance form network $i$ to network $j$ in the NON. (ii) Shell $j$
is a set $L_j$ whose networks are at distance $j$ from the root
network, where $j \in [0,s]$ and $s$ is the total number of shells.
In the following example, we use $i_j$ to denote network $i$ in
shell $j$, e.g., ${i_j} \in L_j$. Note that in shell 0, there is one
and only one network $1_0$. (iii) $g_i(x)$ is the generating
function of network $i$ \cite{Newman2001}, which reflects the
topology of network $i$ and satisfies

\begin{equation}\label{newaddeq1}
g_i(x) = 1- G_{0,i}(x,f_i).
\end{equation}

where $G_{0,i}(x,f_i)$ satisfies
\cite{Newman2001,Newman2002PRE,Sergey2010}
\begin{equation}\label{newaddeq2}
G_{0,i}(x,f_i) =G_{0,i}(xf_i+1-x) =
\sum^{\infty}_{k=0}{P_i(k)(xf_i+1-x)^k},
\end{equation}
and
\begin{equation}\label{newaddeq3}
G_1(x,f)=G^{\prime}_0(x,f)/G^{\prime}_0(1)=f.
\end{equation}

Next, we show analytically the steps in the dynamics of the
cascading failures as demonstrated in Fig. 3.

\begin{figure}[h!]
\centering \includegraphics[width=0.48\textwidth]{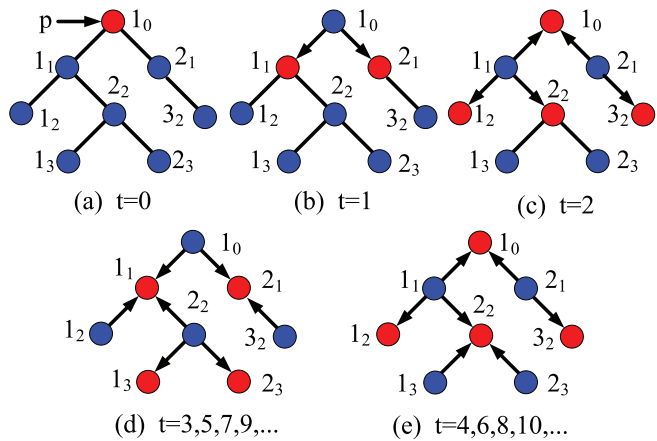}
\caption{The dynamic of cascading failures. In this figure, each
node represents a network. The arrow (on the link) illustrates the
damage spreading from the root network to the whole NON shell by
shell.}\label{figs3}
\end{figure}

Step (0): At $t=0$ (Fig. 3(a)), we begin by randomly removing a
fraction $1-p$ of nodes from the root network (network $1_0$), and
removing all the $A_{1_0}$-links connected to these removed nodes.
Next we remove all the nodes that become disconnected to the largest
component of network $1_0$. Thus at $t=0$, for the network $1_0$,
the fraction of remaining nodes in network $1_0$ after the initial
failure is $x_{0,1_0}=p$ and the fraction nodes in the giant
component $\mu_{0,1_0}=pg_{1_0}(p)$.

Step (1): At $t=1$, the root network spreads its damages to all its
neighboring networks ${i_1} \in L_1$ (Fig. 3(b)). So we remove all
nodes in networks ${i_1}$ that are connected to the removed nodes in
network $1_0$ and then remove all the nodes not in the giant
components of networks ${i_1}$. At $t=1$, the failure of networks
$i_1$ is equivalent to a random removal of the fraction of
$1-x_{1,i_1}$ nodes from networks $i_1$ \cite{Sergey2010}, where
$x_{1,i_1}=pg_{1_0}(x_{0,1_0})$, and the giant component of network
$i_1$ is $\mu_{1,i_1}=x_{1,i_1}g_{i_1}(x_{1,i_1})$.

Step (2): At $t=2$, the networks ${i_1} \in L_1$ reflects their
damages back to the root network and spreads their damages to all
their neighboring networks ${j_2} \in L_2$ (See Fig. 3(c)). So we
remove all nodes in networks $1_0$ and ${j_2}$ that are connected to
the removed nodes in networks ${i_1}$ and then removing all the
nodes not in the giant components of networks ${i_1}$. Again the
failure of network $1_0$ is equivalent to a random removal of the
fraction of $1-x_{2,1_0}$ nodes from networks $1_0$, where
$x_{2,1_0}=p\prod_{i_1 \in L_1}g_{i_1}(x_{1,i_1})$; the failure of
networks $j_2$ is equivalent to a random removal of the fraction of
$1-x_{2,j_2}$ nodes from networks $j_2$, where
$x_{2,j_2}=pg_{1_0}(x_{0,1_0})g_{i_1}(x_{1,i_1})$ for networks $i_1$
that are linked to networks $j_2$.

Step (3): At $t=3$, the root network spreads its further damages to
the networks ${i_1}$ in shell 1 again, the networks ${j_2}$ in shell
2 reflect their damages back to the neighboring networks ${i_1}$ in
shell 1, and to the neighboring networks ${u_3}$ in shell 3 as shown
in Fig. 3(d). A network in ${i_1}$ receives the damages information
$x_{2,1_0}$ from network $1_0$, $x_{2,i_2}$ from networks ${j_2}$
that are linked to networks $i_1$, and $x_{1,v_1}$ from networks
${v_1}$ in shell 1 where the networks $v_1$ are the neighboring
networks of $i_1$'s neighboring networks, i.e., the distance between
networks $i_1$ and networks $v_1$ is 2. Thus we can obtain that
$x_{3,i_1}=pg_{1_0}(x_{2,1_0})\prod_{j_2}g_{j_2}(x_{2,j_2})\prod_{v_1}g_{v_1}(x_{1,v_1})$.
Similarly, we obtain the failure of networks ${u_3} \in L_3$ to be
$x_{3,u_3}=pg_{1_0}(x_{0,1_0})g_{i_1}(x_{1,i_1})g_{j_2}(x_{2,j_2})$,
where networks ${u_3}$ in shell 3 are connected to networks ${j_2}$
in shell 2 and networks ${j_2}$ are connected to networks ${i_1}$ in
shell 1.

\begin{figure}[h!]
\centering \includegraphics[width=0.48\textwidth]{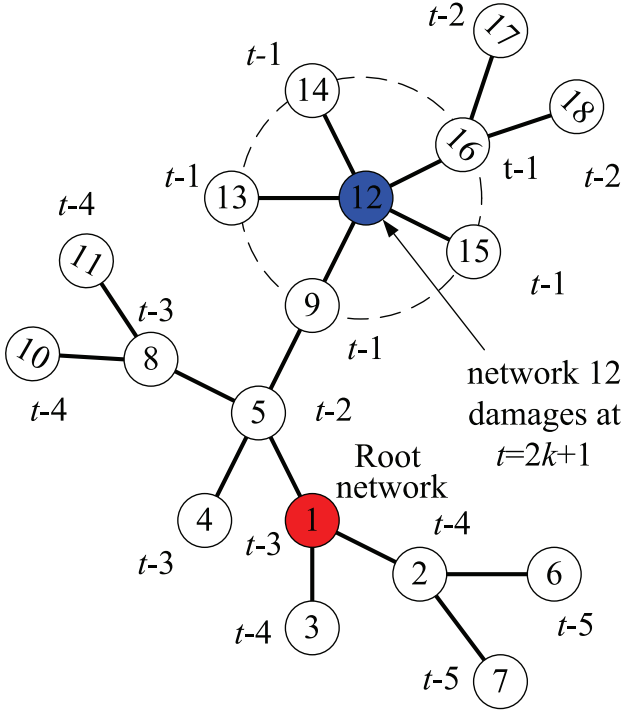}
\caption{How does the damage spread in a NON system? In this figure,
each node represents a network. When looking at network 12 for
example, it becomes damaged at $t=2k+1$ ($k=1,2,3,...$). It receives
the damage from network 8 at $t=2k+3$, because network 8 gets damage
at $t=2$ for the first time and its damage spreads to network 12
when $t=5$ for the first time, which agrees with Eqs.
(\ref{eq2}-\ref{eq6}) that network $i$ receives damage from network
$j$ if and only if $t-D_{ij} \geq D_{1j}$.}\label{figs4}
\end{figure}

We continue the cascading process step by step [see Figs. 3(d) and
(e)] until the convergence step, $t=\tau$, when no further nodes and
links removal occurs. Accordingly, we investigate the dynamically
cascading process of our model of the loopless NON. First we
initialize the NON as

\begin{equation}\label{eq2}
x_{0,1_0}=p, \mu_{0,1}=pg_{1_0}(p),
\end{equation}
and
\begin{equation}\label{eq3}
g_{i_j}(x_{t,i_j})=\mu_{t,i_j}=1, t < j.
\end{equation}

Thus we can obtain that the giant component of network $i_j$ in
shell $j$, $\mu_{t,i_j}$ at step $t$ satisfies
\begin{equation}\label{eq4}
\mu_{t,i_j}=x_{t,i_j}g_{i_j}(x_{t,i_j}), j \in C(t),
\end{equation}
\begin{equation}\label{eq5}
\mu_{t,i_j}=\mu_{t-1,i_j}, j \notin C(t).
\end{equation}
where $C(t)$ satisfies the sequence

$\left\{
\begin{array}{lcl}
C(0)=\{0\} & \mbox{} &  \\
C(1)=\{1\}& \mbox{} &  \\
C(2)=\{0,2\} & \mbox{} &  \\
C(3)=\{1,3\}& \mbox{} &  \\
C(4)=\{0,2,4\} & \mbox{} &  \\
C(5)=\{1,3,5\}& \mbox{} &  \\
C(6)=\{0,2,4,6\} & \mbox{} &  \\
C(7)=\{1,3,5,7\}& \mbox{} &  \\
C(8)=\{0,2,4,6,8\}& \mbox{} &  \\
...& \mbox{} &  \\
\end{array}\right.$,

and $x_{t,i}$ satisfies
\begin{equation}\label{eq6}
x_{t,i}=p\prod_{j=1,j \neq i}^n g_j(x_{t-D_{ij},i}).
\end{equation}
Furthermore, when $t \longrightarrow \infty$, $\mu_i \equiv
\mu_{\infty}$ and $x_{t-D_{ij},i} = x_{t,i} \equiv x_{i}$. So from
Eqs. (\ref{eq2}-\ref{eq6}) we can obtain that

\begin{equation}\label{eq7}
x_i = p\prod_{j=1,j \neq i}^n g_j(x_j),
\end{equation}
and
\begin{equation}\label{eq8}
\mu_{\infty} = p\prod_{j=1}^n g_j(x_j).
\end{equation}

We also demonstrate how does the damage spread in another example of
NON shown in Fig. 4.

In Figs. 5 and 6 we compare our theoretical results Eqs. (\ref{eq7})
and (\ref{eq8}) with simulation results for 3 different types of
networks, ER networks, RR networks and SF networks. We find that
while the dynamics is different for the three topologies shown in
Fig. 2, the final $P_{\infty} \equiv \mu_{\infty,1}$ is the same as
predicted by the theoretical results, Eq. (\ref{eq22}).

Next, we study the case for $n$ coupled networks where all networks
are with the same degree distribution specified by the generating
functions $G_0(x,f)$ (Eq. (\ref{newaddeq2})) and $G_1(x,f)$ (Eq.
\ref{newaddeq3})). By substituting Eq. (\ref{newaddeq1}) and
(\ref{newaddeq2}) into the Eq. (\ref{eq7}) and introducing the
parameter $z=xf+1-x$, we obtain

\begin{equation}\label{newaddeq4}
x = p(1-G_0(z))^{n-1}
\end{equation}

From Eqs. (\ref{newaddeq3}), (\ref{newaddeq4}) and (\ref{eq8}), the
equations for mutual giant component become

\begin{equation}\label{addeq1}
\frac{1}{p}=\frac{(1-G_0(z))^{n-1}(1-G_1(z))}{1-z},
\end{equation}
and
\begin{equation}\label{addeq2}
P_\infty=p(1-G_0(z))^{n}= \frac{(1-G_0(z))(1-z)}{1-G_1(z)}.
\end{equation}

\begin{figure}[h!]
\centering \includegraphics[width=0.23\textwidth]{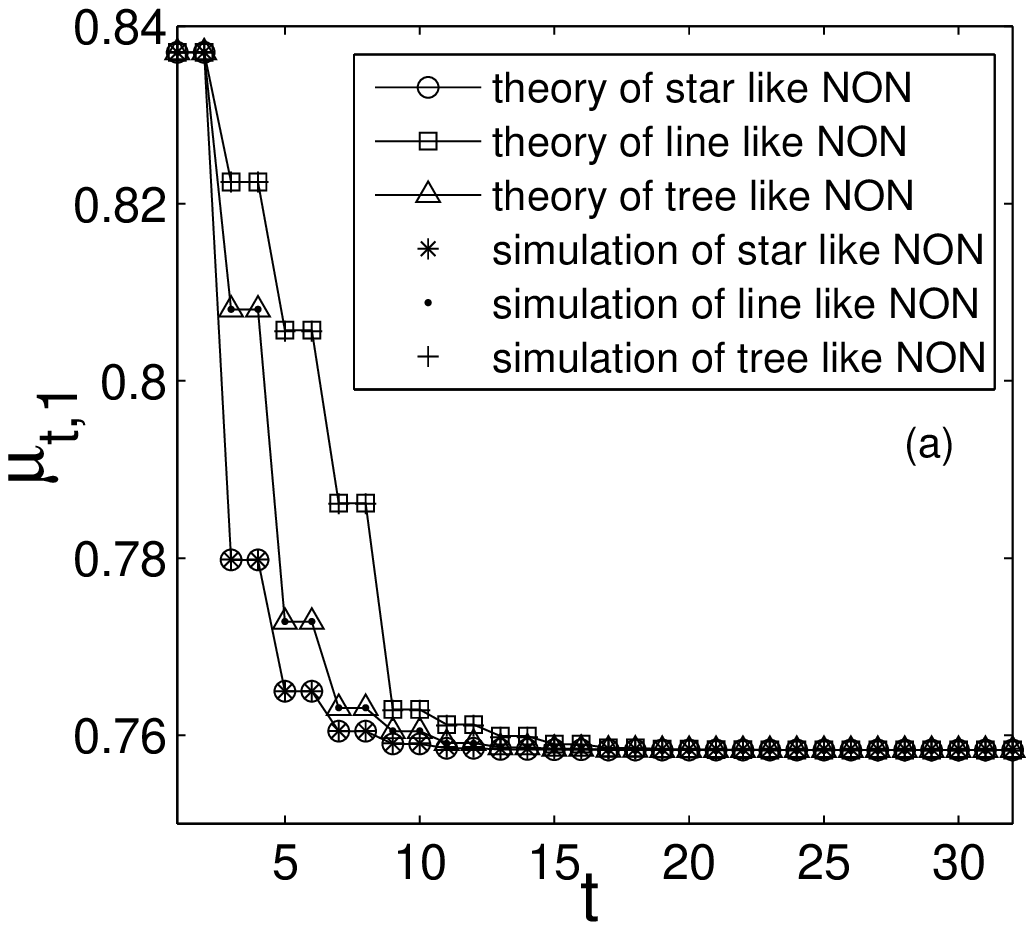}
\hspace{.0in} \includegraphics[width=0.23\textwidth]{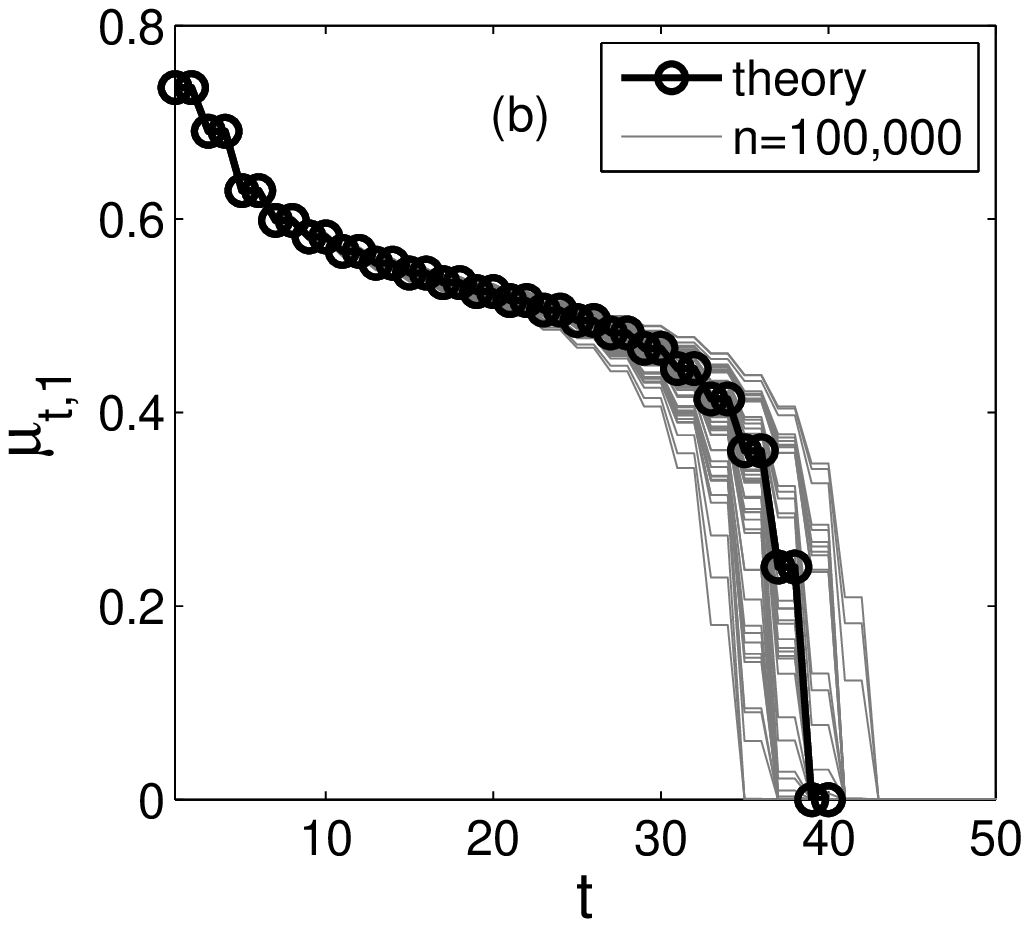}
\caption{(a) Simulation results of the giant component of the root
network $\mu_{t,1}$ after $t$ cascading failures for three types of
NON composed of 5 ER networks shown in Fig.~2. For each network in
the NON, $N=100,000$ and $k = 5$. The value of $p$ chosen is
$p=0.85$, and the predicated threshold $p_c=0.76449$ (from Eqs. (22)
and (24)). All points are the results of averaging over 40
realizations. Note that while the dynamics is different for the
three topologies, the final $P_{\infty} \equiv \mu_{\infty,1}$ is
the same, i.e., the final $P_{\infty}$ does not depend on the
topology of the NON. (b) Simulations of the giant component,
$\mu_{t,1}$, for the tree-like NON [Fig. 2] with the same parameters
as in (a) but for $p=0.755<p_c=0.76449$. The figure shows 50
simulated realizations of the giant component left after $t$ stages
of the cascading failures compared with the theoretical prediction
of Eqs. (\ref{eq2})-(\ref{eq6}).}\label{fig5}
\end{figure}

\begin{figure}[h!]
\centering \includegraphics[width=0.23\textwidth]{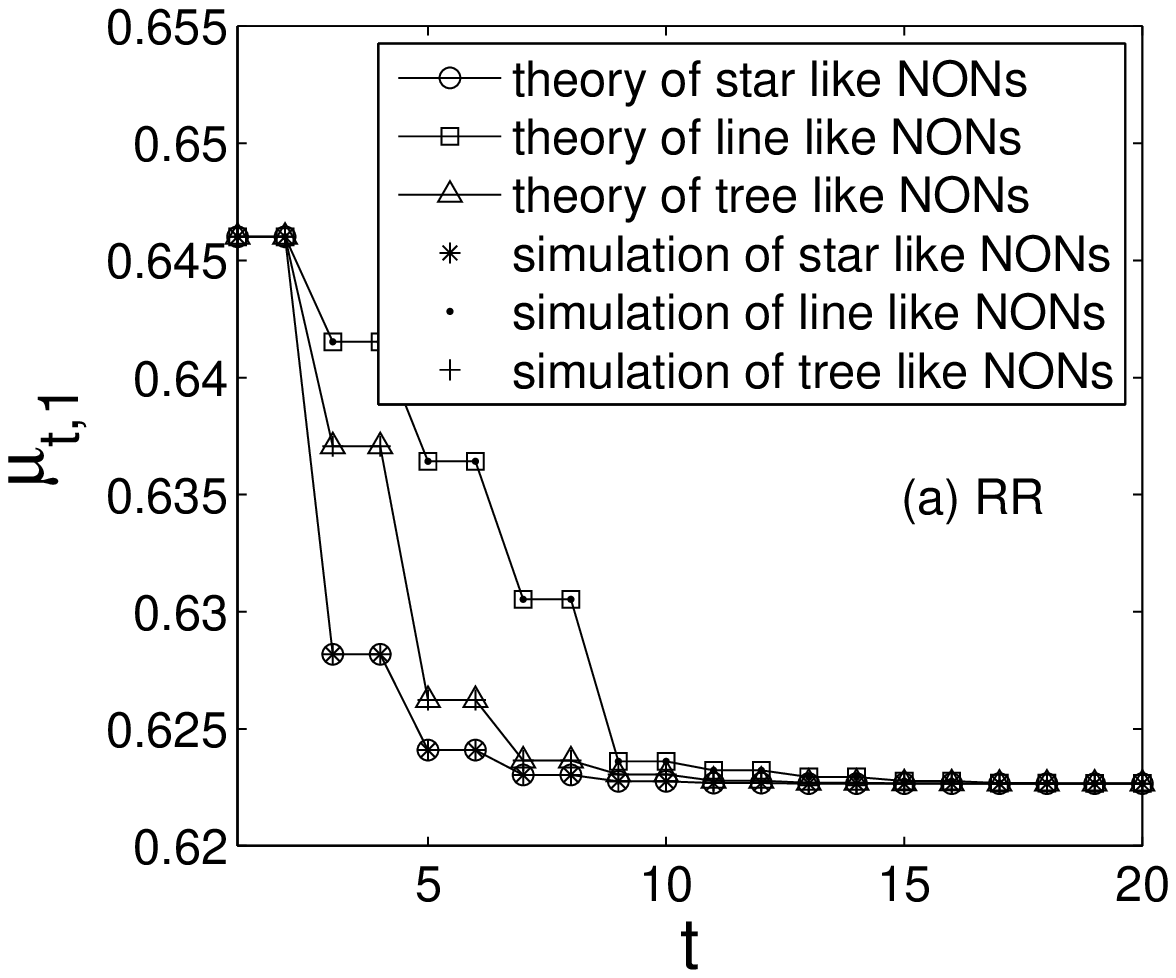}
\centering \includegraphics[width=0.23\textwidth]{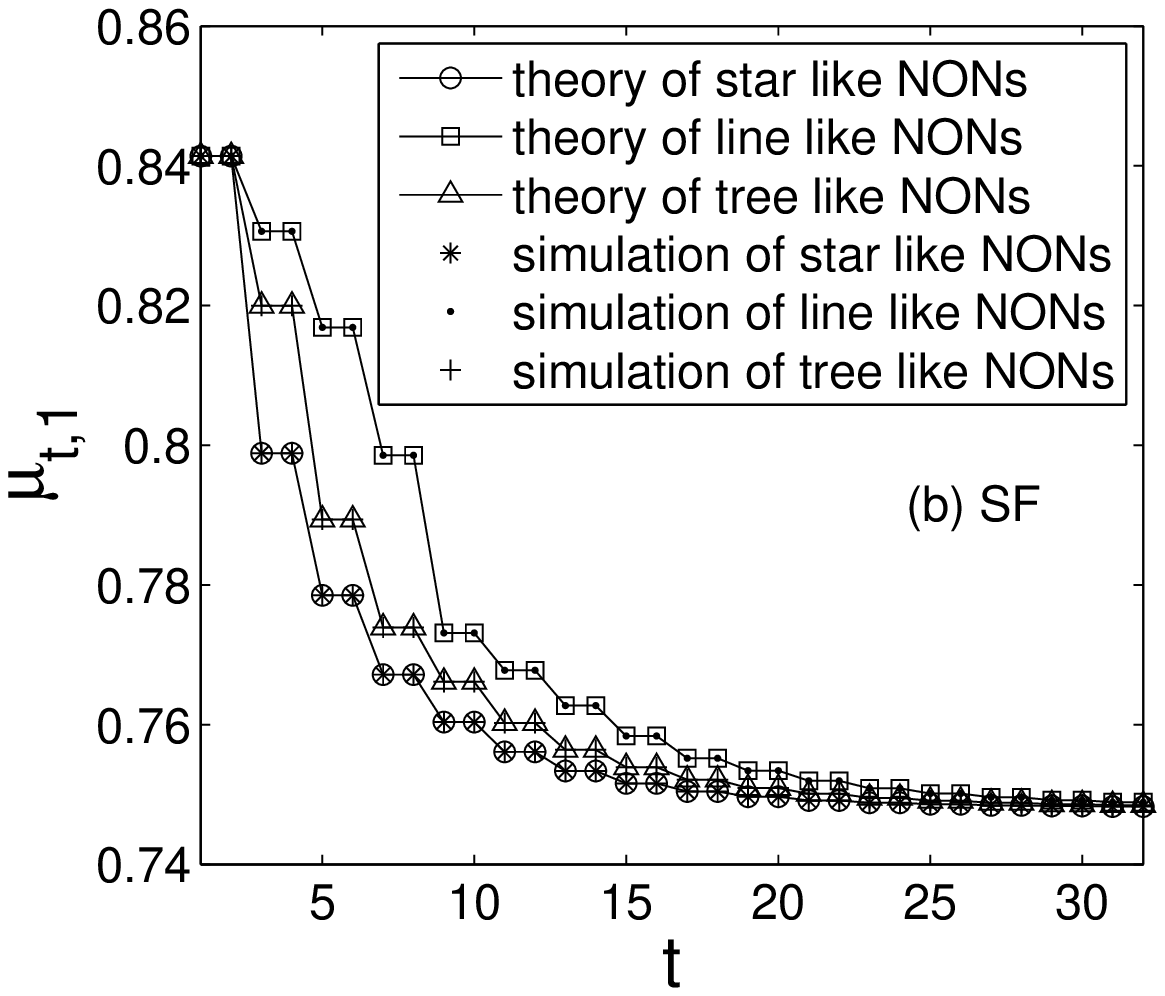}
\caption{(a) Simulation results for the giant component of the root
network $\mu_{t,1}$ after $t$ cascading failures for three types of
NON composed of 5 RR networks. The structures of the NON are as
shown in Fig.~2. For each network in the NON, $N=100,000$ and $k=5$
. The value of $p$ chosen is $p=0.65$, and the predicated threshold
$p_c=0.6047$ [from Eqs. (\ref{eq36}) and (\ref{eq38})]. The points
are the results of averaging over 40 realizations. It is seen that
while the dynamics is different for the three topologies, the final
$P_{\infty} \equiv \mu_{\infty,1}$ is the same, i.e., the final
$P_{\infty}$ does not depend on the topology of the NON. (b)
Simulation results of the giant component of the root network
$\mu_{t,1}$ after $t$ cascading failures for three types of NON
composed of 5 SF networks shown in Fig.~2. For each network in the
NON, $N=100,000$ and $\lambda = 2.3$ . The value of $p$ chosen is
$p=0.85$(above $p_c$). The points are the results of averaging over
40 realizations. We also can see here that while the dynamics is
different for the three topologies, the final $P_{\infty} \equiv
\mu_{\infty,1}$ is the same.}\label{fig6}
\end{figure}

\section{THE CASE of NON COMPOSED OF $n$ ER NETWORKS}

The case of NON of $n$ Erd\H{o}s-R\'{e}nyi
(ER)~\cite{ER1959,ER1960,Bollob1985} networks with average degrees
$k_1, k_2,... k_i,...,k_n$ can be solved explicitly \cite{gao2010}.
In this case, the generating functions of the $n$ networks are
\cite{Newman2002PRE}.
\begin{equation}\label{eq9}
G_{1,i}(x)=G_{0,i}(x)=\exp[k_i(x-1)].
\end{equation}
Accordingly, we obtain that the generating function $g_i(x_i)$
satisfies
\begin{equation}\label{eq10}
g_i(x_i) =1-\exp[k_i x_i(f_i-1)],
\end{equation}
where $f_i=\exp[k_i x_i(f_i-1)]$ and thus $g_i(x_i)=1- f_i$. Using
Eq. (\ref{eq7}) for $x_i$ we get
\begin{equation}\label{eq11}
f_{i} = \exp[-pk_i \prod_{j=1}^{n} (1-f_j)], i = 1,2,...,n.
\end{equation}

By introducing a new variable $r=f_i^{1/k_i},i=1,2,...,n$ into Eq.
(\ref{eq11}), we can reduce the $n$ equations to a single equation,

\begin{equation}\label{eq12}
r = \exp[-p \prod\limits_{i=1}^{n} (1-r^{k_i})],
\end{equation}
which can be solved graphically for any $p$. For small $p$, Eq.
(\ref{eq12}) has only the trivial solution $r=1$. This case
corresponds to the absence of the mutual giant component and hence
to the complete fragmentation of the networks. As $p$ increases a
nontrivial solution $r<1$ emerges at some critical value of $p=p_c$.
The critical case corresponds to the tangential condition:
\begin{equation}\label{eq13}
1 = \frac{d}{dr}\exp[-p \prod\limits_{i=1}^{n} (1-r^{k_i})].
\end{equation}

Thus, the critical value of $r$ satisfies a transcendental equation
\begin{equation}\label{eq14}
r = \exp \Big \{-\frac{\prod\limits_{i=1}^{n}
  (1-r^{k_i})}{\sum\limits_{i=1}^{n}{[k_ir^{k_i} \prod\limits_{j=1,j
        \neq i}^{n}(1-r^{k_j})]}} \Big \}.
\end{equation}

From Eqs. (\ref{eq11}) and (\ref{eq13}) we can obtain the critical
percolation threshold $p_c$ and the the value of $\mu_{\infty}$ at
$p_c$ as
\begin{equation}\label{eq15}
p_c = \Big \{\sum\limits_{i=1}^{n}{\big [k_if_i \prod\limits_{j=1,j
\neq i}^{n}(1-f_j) \big ]} \Big \}^{-1},
\end{equation}
and
\begin{equation}\label{eq16}
\mu_{\infty}=p\prod\limits_{i=1}^{n}(1-f_i).
\end{equation}

If $p<p_c$, Eqs. (\ref{eq11}) have only the trivial solutions
($f_i=1$) and $\mu_{\infty} \equiv 0$. When the $n$ networks have
the same average degree $k$, $k_i=k$ ($i=1,2,...,n$), we obtain from
Eq. (\ref{eq11}) that $f_c \equiv f_i(p_c)$ satisfies \cite{gao2010}
\begin{equation}\label{eq17}
f_c = e^{\frac{f_c-1}{nf_c}}.
\end{equation}
This solution can be expressed in terms of the Lambert function
$W(x)$~\cite{Lambert1758,Corless1996},
\begin{equation}\label{eq18}
f_c = -[nW(-\frac{1}{n}e^{-\frac{1}{n}})]^{-1}.
\end{equation}
Once $f_c$ is known, we obtain $p_c$ and $\mu_{\infty,n} \equiv
P_{\infty}$ at $p_c$ by substituting $k_i=k$ into Eqs. (\ref{eq15})
and (\ref{eq16})

\begin{equation}\label{eq19}
p_c = [nkf_c(1-f_c)^{(n-1)}]^{-1}
\end{equation}
and
\begin{equation}\label{eq20}
P_{\infty} =\frac{1-f_c}{nkf_c}.
\end{equation}

For $n=1$ we obtain the known results $p_c=1/k$ and $P_{\infty}=0$
at $p_c$ (representing the second order transition) of
Erd\H{o}s-R\'{e}nyi \cite{ER1959,ER1960,Bollob1985}. Substituting
$n=2$ in Eqs. (\ref{eq19}) and (\ref{eq20}) one obtains the exact
results derived in \cite{Sergey2010}. Note that for all $n>1$ we
obtain $P_{\infty}
> 0$ at $p_c$ representing a first order nature of the percolation
transition. For the behavior of $p_c$ [Eq.(24)] for large $n$ see
Appendix A.

\subsection{The minimum degree $k$ and the giant component $P_{\infty}(p)$}

\begin{figure}[h!]
\centering \includegraphics[width=0.23\textwidth]{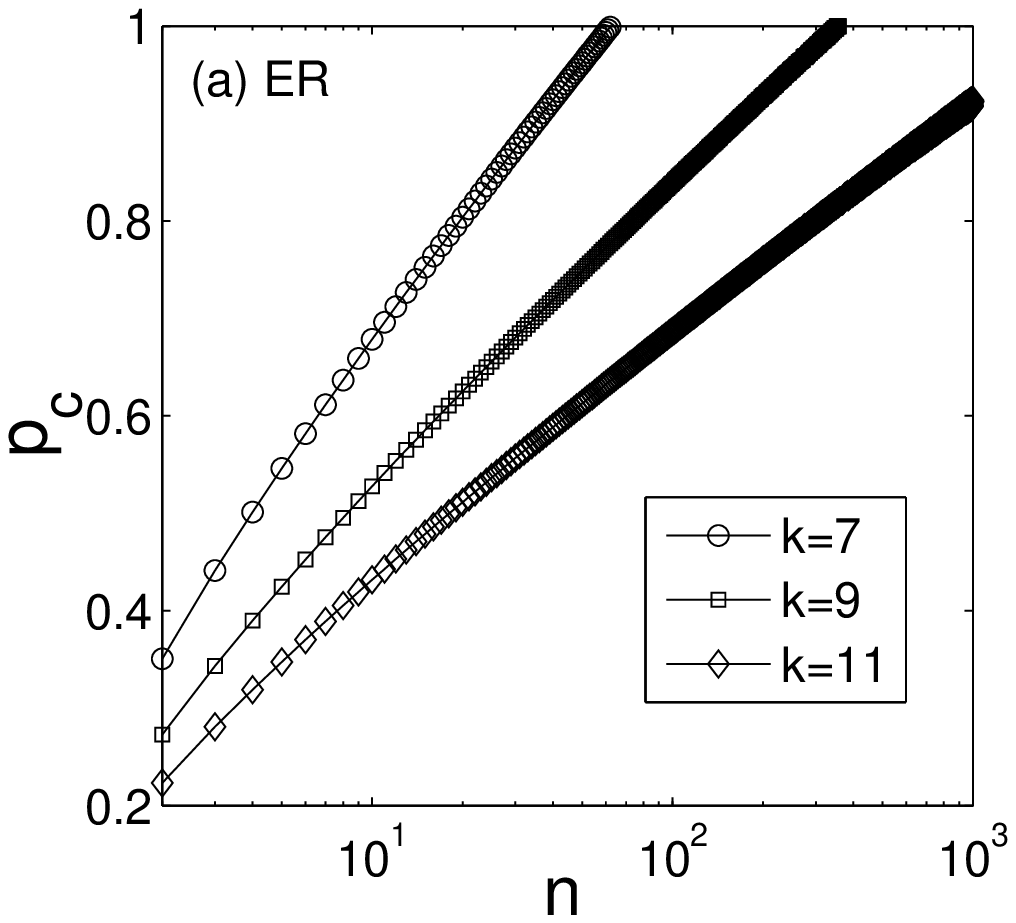}
\hspace{.0in}\includegraphics[width=0.23\textwidth]{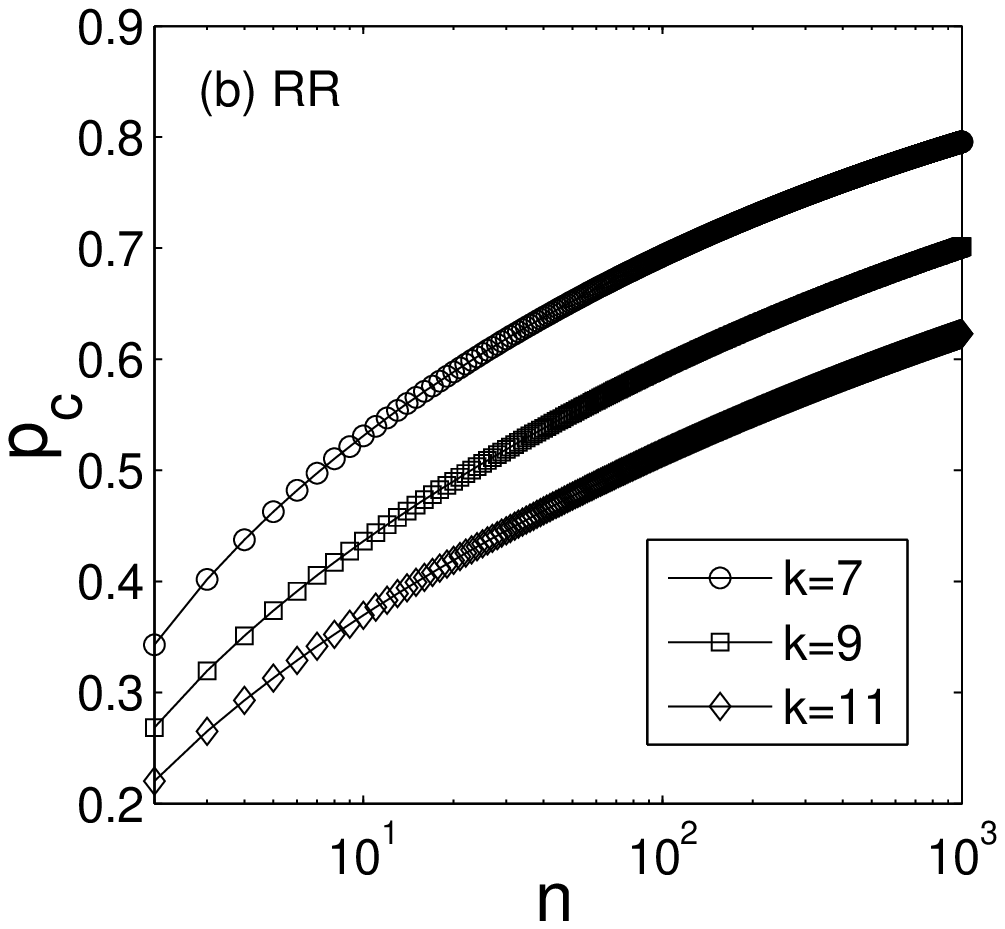}
\caption{The critical fraction $p_c$ for different $k$ and $n$ for
(a) ER NON system and (b) RR NON system. The results for ER NON
system are obtained by Eqs. (\ref{eq17}) and (\ref{eq19}), while the
results of RR NON system are obtained from the solution of Eqs.
(\ref{eq36}) and (\ref{eq38}). The results are in good agreement
with simulations. In the simulations $p_c$ was calculated from the
number of cascading failures which diverge at $p_c$
\protect\cite{parshani2011}}
\end{figure}\label{fig7}

Here we show that while for the ER NON system there exists a minimum
degree $k_{\min}(n)$ below which the NON collapses, in the RR NON
such a $k_{\min}$ does not exists and the RR NON is stable
($p_c(n)<1$) for any $n$. Thus, the RR NON is significantly more
robust compared to the ER NON due to the critical role of the singly
connected nodes on the vulnerability of the NON system. To analyze
$p_c$ as a function of $n$, we find $f_c$ from Eq. (\ref{eq17}) and
substitute it into Eq. (\ref{eq19}), and we obtain $p_c$ as a
function of $n$ for different $k$ values, as shown in Fig.~7(a) for
the ER NON system. It is seen that the NON becomes more vulnerable
with increasing $n$ or decreasing $k$ ($p_c$ increases when $n$
increases or $k$ decreases). Furthermore, for a fixed $n$, when $k$
is smaller than a critical number $k_{min}(n)$, $p_c \geq 1$ meaning
that for $k<k_{min}(n)$, the NON will collapse even if a single node
fails. From Eq. (\ref{eq19}) by substituting $p_c=1$, we get the
minimum of $k$ as a function of $n$

\begin{equation}\label{eq21}
k_{\min}(n) = [nf_c(1-f_c)^{(n-1)}]^{-1}
\end{equation}

Note that Eq. (\ref{eq21}) together with Eq. (\ref{eq17}) yield the
value of $k_{\min}(1)=1$ for $n=1$, reproducing the known ER result,
that $\langle k \rangle =1$ is the minimum average degree needed to
have a giant component. For $n=2$, Eq. (\ref{eq21}) yields the
result obtained in \cite{Sergey2010}, i.e., $k_{\min}=2.4554$.

When the $n$ networks have the same average degree $k$, $k_i=k$
($i=1,2,...,n$), using Eqs. (17) and (21) we obtain the percolation
law for the order parameter, the size of the mutual giant components
for all $p$ values and for all $k$ and $n$ \cite{gao2010},
\begin{equation}\label{eq22}
\mu_{\infty,n} \equiv P_{\infty}=p[1-\exp(-kP_{\infty})]^n.
\end{equation}
The solutions of equation (\ref{eq22}) for several $n$-values are
shown in Fig. 8(a). Results are in excellent agreement with
simulations. The special case $n=1$ is the known ER second order
percolation law for a single network
\cite{ER1959,ER1960,Bollob1985}.

\subsection{The case of NON with different average degrees}

Next, we study the case where the average degrees of all $n$
networks is not the same. Without loss of generality we assume that
$m$ networks have the same average degree $\langle k\rangle _2$, and
other $n-m$ networks have the same average degree $\langle k\rangle
_1$. We define $\alpha=\langle k\rangle_1/\langle k\rangle _2$ where
$0<\alpha \leq 1$. Using Eqs. (\ref{eq11}-\ref{eq14}) we can show
that $f_c$ satisfies
\begin{equation}\label{eq23}
f_c=\exp[\frac{(f_c-1)(1-f_c^{1/\alpha})}{(n-m)f_c(1-f_c^{1/\alpha})+
    mf_c^{1/\alpha}(1-f_c)/\alpha}].
\end{equation}
Results for $p_c$ and the mutual giant component for different
values of m, and alpha are shown in Fig.~11. The case of $\langle
k\rangle_1 \ll \langle k\rangle _2$ is interesting, since in this
limit the $m$ networks with large $\langle k\rangle_2$ due to their
good connectivity can not cause further damage to the $n-m$ networks
with $\langle k\rangle_1$. Thus the NONs system can be regarded as
only $n-m$ networks. Indeed, when $\alpha\rightarrow 0$, $f_c$
satisfies
\begin{equation}\label{eq24}
f_c = e^{\frac{f_c-1}{(n-m)f_c}}.
\end{equation}
And then equation of $p_c$ and $\mu_{\infty}$ are obtained as
\begin{equation}\label{eq25}
p_c = [(n-m) \langle k\rangle_1f_c(1-f_c)^{(n-m-1)}]^{-1},
\end{equation}
and
\begin{equation}\label{eq26}
\mu_{\infty}=\frac{1-f_c}{(n-m)\langle k\rangle_1f_c}.
\end{equation}
Equations (\ref{eq24}-\ref{eq26}) are indeed the same as Eqs.
(\ref{eq17}, \ref{eq19}, and \ref{eq20}) where $n$ is replaced by
$n-m$. This result is seen also in Fig.~11, where the limit of
$\alpha=0$ yield the same results as for $\alpha=1$ for $n-m$
networks.

When $\langle k\rangle_1 \ll \langle k\rangle_2$ for any $p$, we can
get the equation of $P_{\infty}$ as a function of, $p$, $m$, $n$ and
$\langle k\rangle_1$
\begin{equation}\label{eq27}
P_{\infty} = p[1-\exp(-\langle k\rangle_1P_{\infty})]^{n-m}.
\end{equation}
The average number of cascading stages, $\langle \tau \rangle$, as a
function of $p$ for different value of average degree is shown in
Fig.~11. The numerical simulation results show that $\tau$ increases
sharply when $p$ is near $p_c$ \cite{parshani2010}.

\begin{figure}[h!]
\centering
\includegraphics[width=0.23\textwidth]{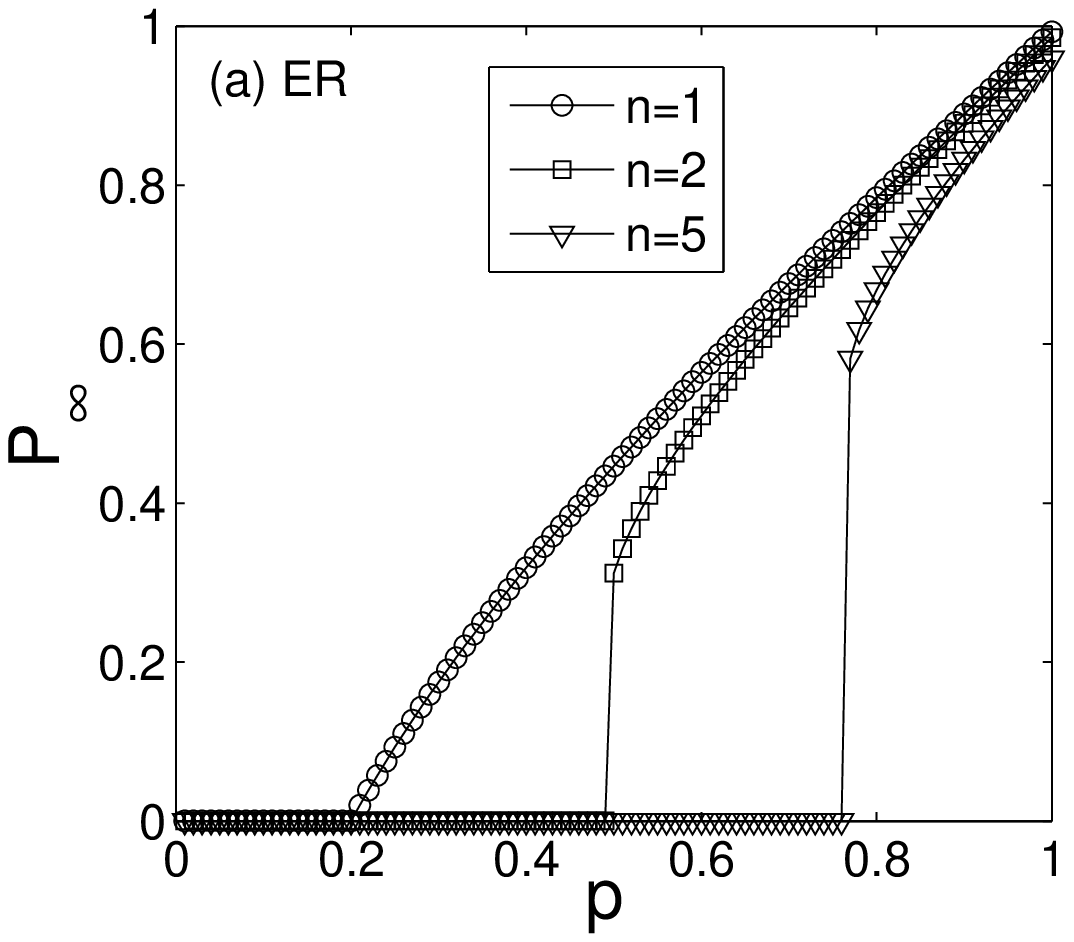}
\centering
\includegraphics[width=0.23\textwidth]{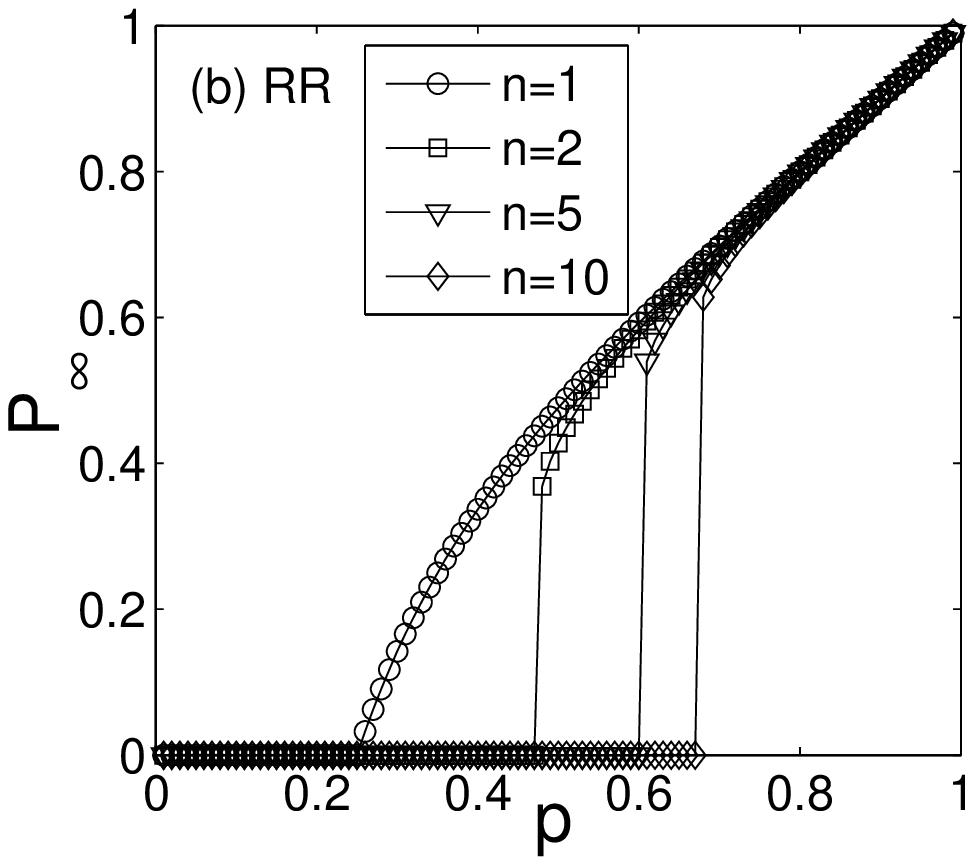}
\centering
\includegraphics[width=0.23\textwidth]{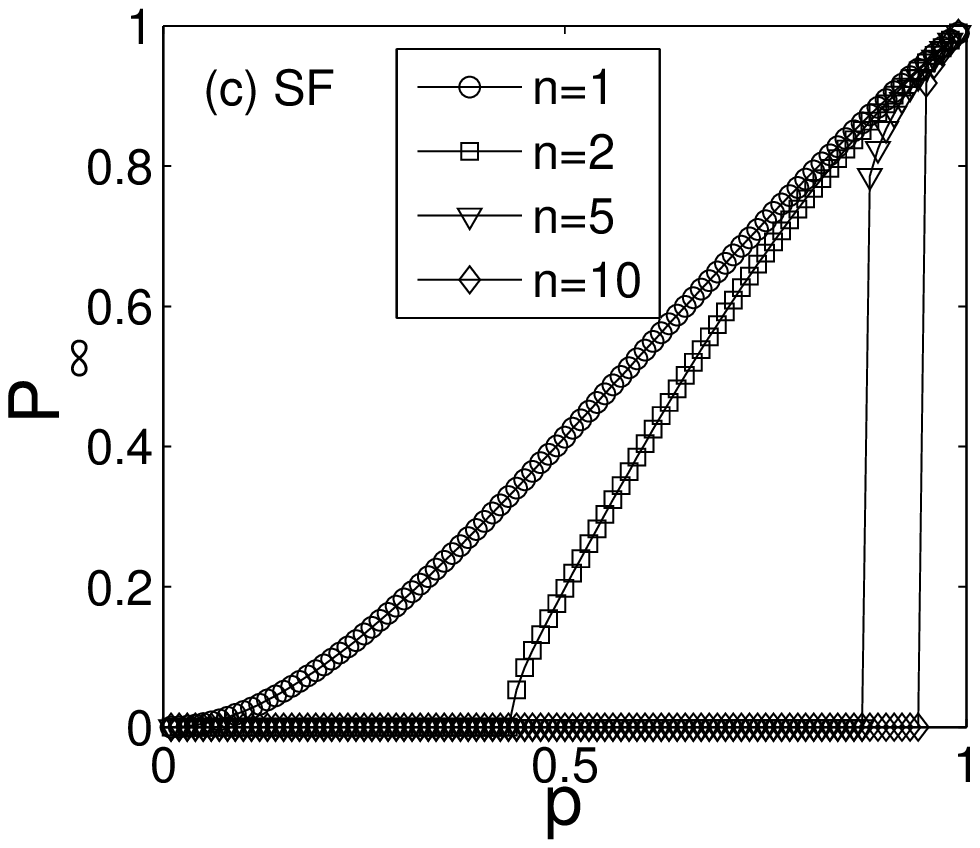}
\caption{ Loopless NON is composed of (a) ER networks, (b) RR
networks and (c) SF networks. Plotted is $P_{\infty}$ as a function
of $p$ for $k=5$ (for ER and RR networks) and $\lambda=2.3$ for SF
networks and several values of $n$. The results obtained using Eq.
(\ref{eq22}) for ER networks, Eq. (40) for RR networks and Eq. (51)
for SF networks, agree well with simulations.}\label{fig8}
\end{figure}

\begin{figure}[h!]
\centering\includegraphics[width=0.23\textwidth]{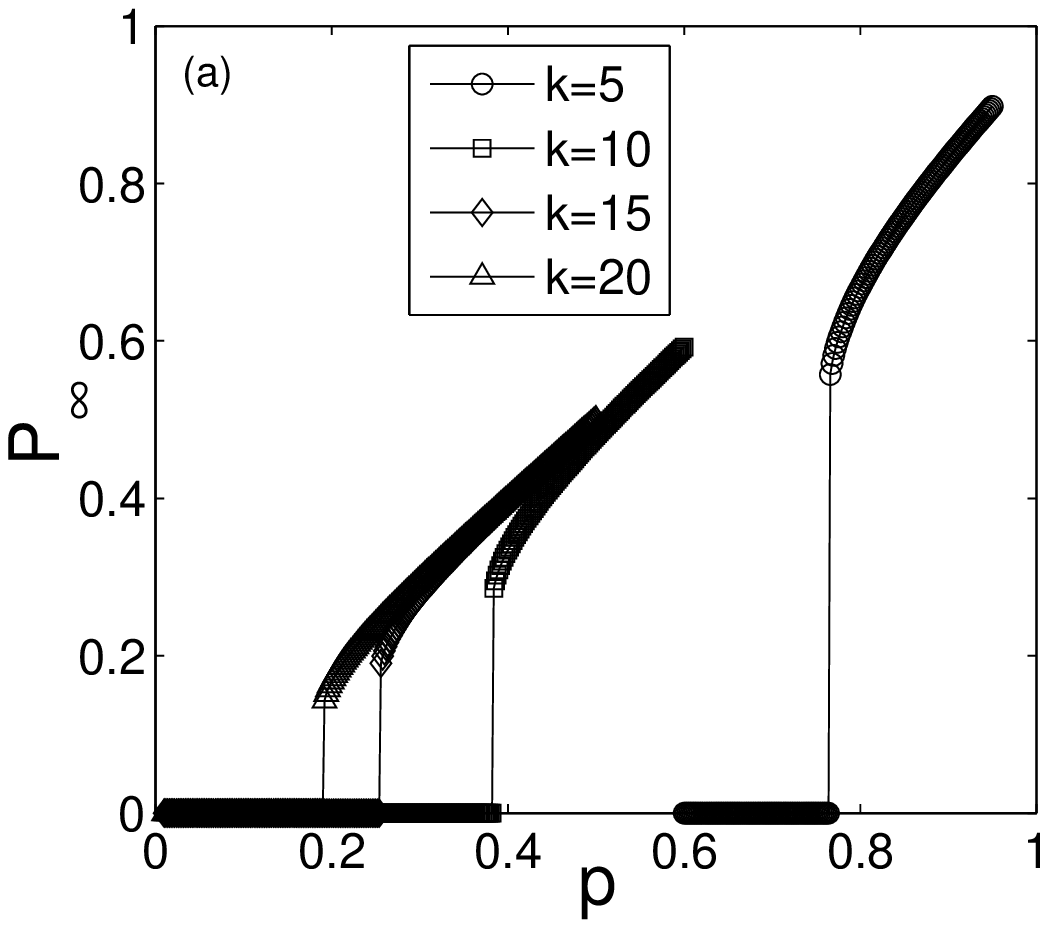}
\centering\includegraphics[width=0.23\textwidth]{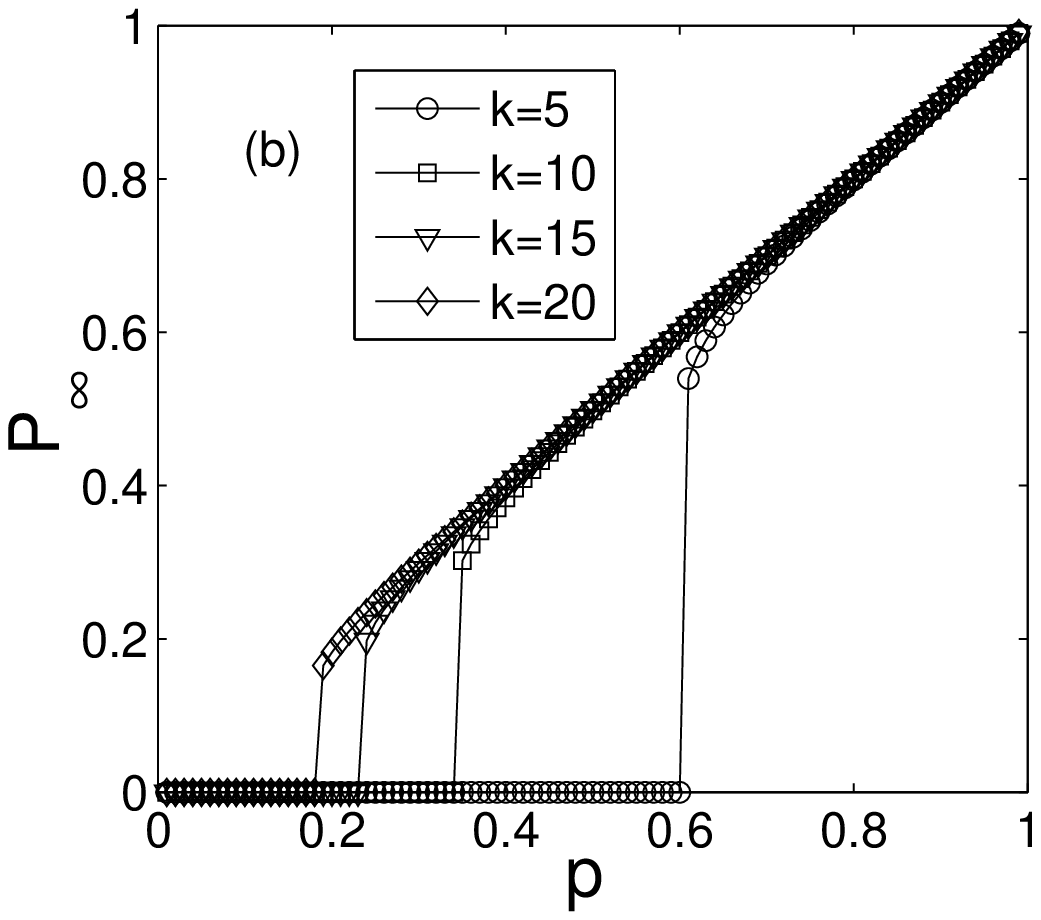}
\centering\includegraphics[width=0.23\textwidth]{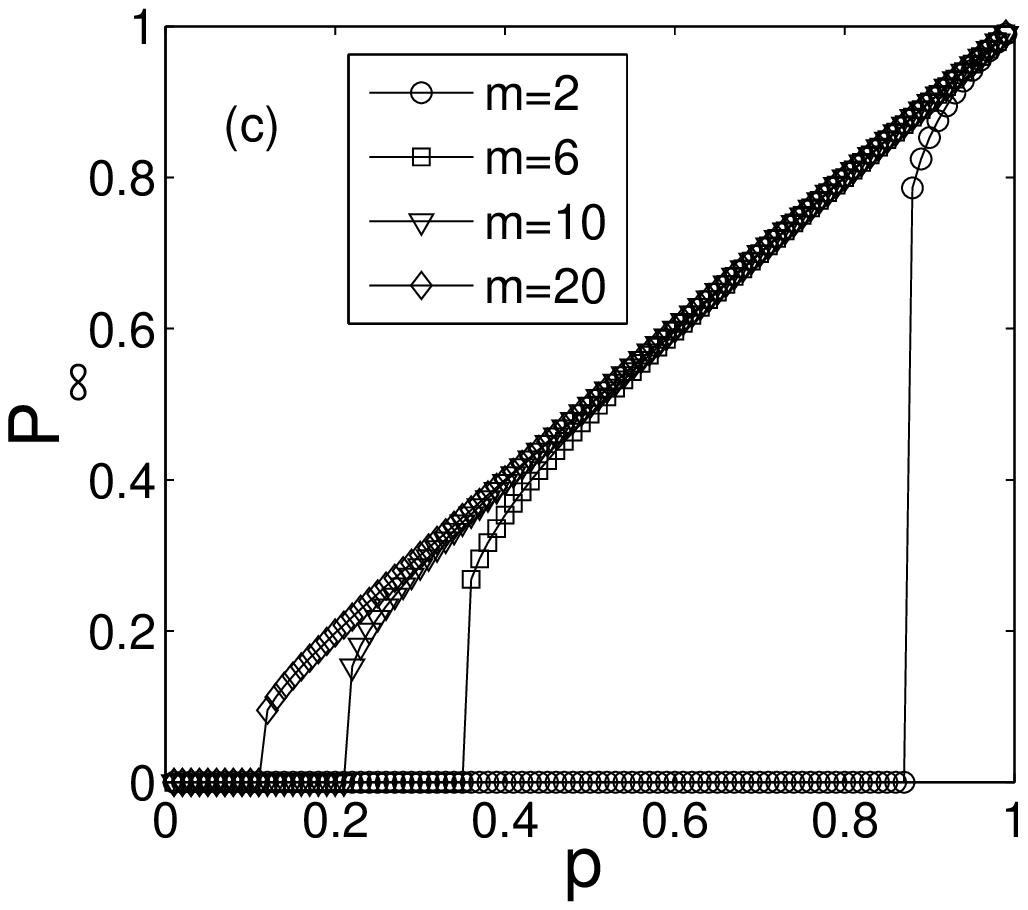}
\caption{ Loopless NON is composed of (a) ER networks, (b) RR
networks and (c) SF networks. Plotted is $P_{\infty}$ as a function
of $p$ for $n=5$ for several values of $k$ (ER and RR networks) and
several values of $m$ (SF networks for $\lambda=2.3$). The results
obtained using Eq. (\ref{eq22}) for ER networks, Eq. (\ref{eq40})
for RR networks and Eq. (\ref{eq51}) for SF networks, agree well
with simulations. }\label{fig9}
\end{figure}

\begin{figure}[h!]
\centering \includegraphics[width=0.44\textwidth]{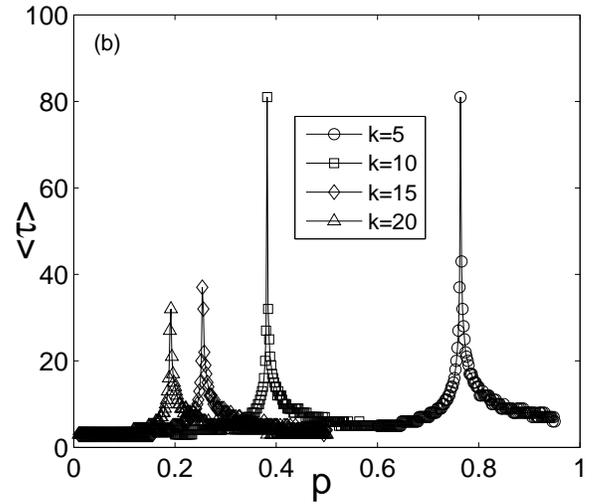}
\centering \caption{For star-like network of 5 ER networks, the
average convergence stage $\langle \tau\rangle$ as a function of $p$
for different $k$. In the simulation, $N=10^6$, and the simulation
results are obtained by over 30 realizations. This feature enables
to find accurate estimate for $p_c$ in simulations
\protect\cite{parshani2010}.) agree well with
simulations.}\label{fig10}
\end{figure}

\begin{figure}[h!]
\centering
\includegraphics[width=0.23\textwidth]{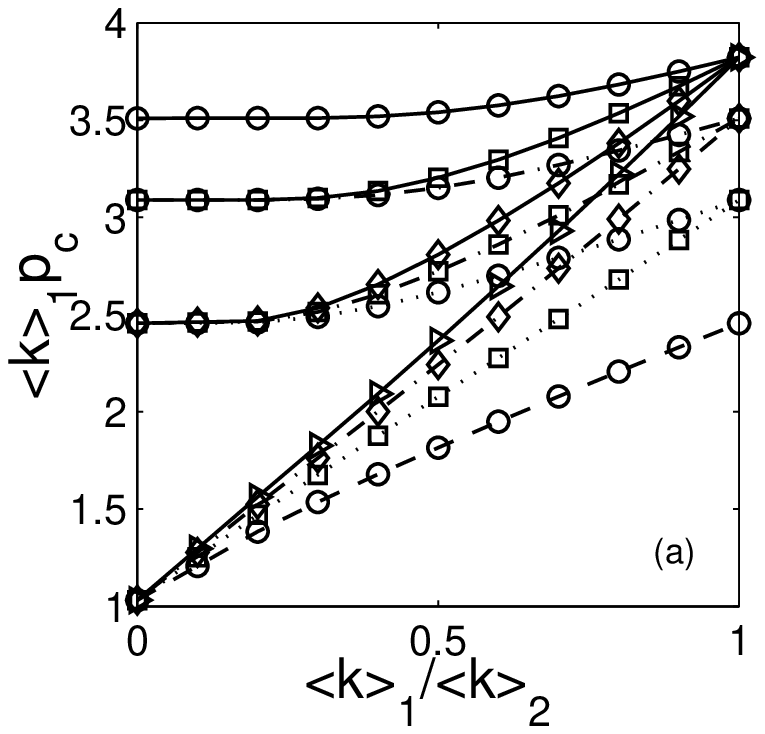}
\hspace{.0in}
\includegraphics[width=0.23\textwidth]{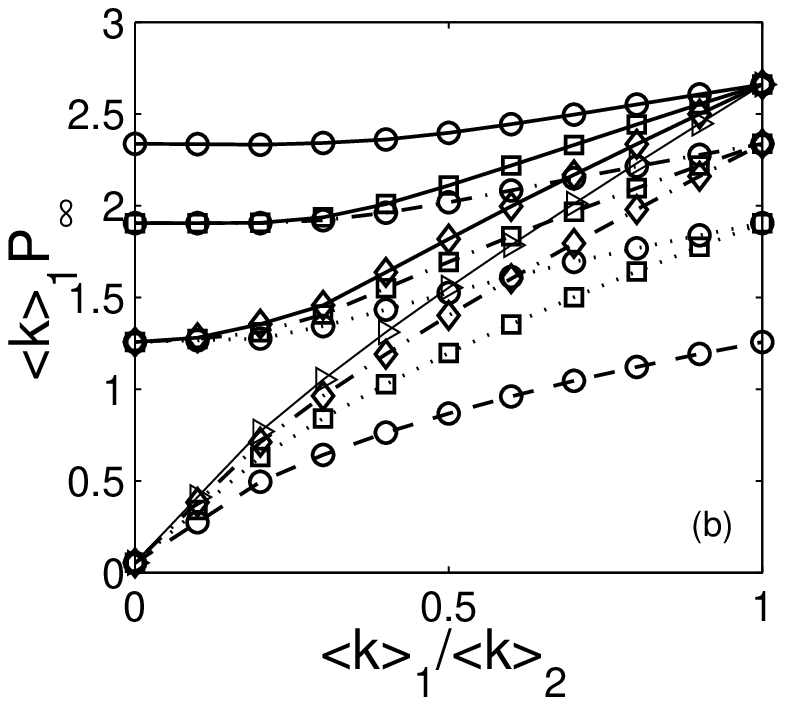}
\caption{ For loopless network of $n$ ER networks, $\langle
k\rangle_1p_c$ and $\langle k\rangle_1\mu(\infty)$ as function of
the ratio $\langle k\rangle_1/\langle k\rangle_2$ for $n=2$
(dashed), $n=3$ (dotted), $n=4$ (dashdot) and $n=5$ (solid) and for
$m=1$ ($\circ$), $m=2$ ($\Box$), $m=3$ ($\diamond$) and $m=4$
($\triangleright$), where $m$ denotes the number of individual
networks whose average degree are the same $\langle k\rangle_2$ and
average degree of the other $n-m$ networks are $\langle k\rangle_1$.
The results are obtained using
Eqs.(\ref{eq24})-(\ref{eq26}).}\label{fig11}
\end{figure}

\section{ANALYTICAL RESULTS FOR THE CASE OF NON OF $n$ RR NETWORKS}

Next, we study the case for a tree-like NON of $n$ RR networks. The
degree of network $i$ is $k_i$. The generating functions of network
$i$ are
\begin{equation}\label{eq28}
G_{i,0}(x)=\sum k_{i}x^{k_i}=x^{k_i},
\end{equation}
and
\begin{equation}\label{eq29}
G_{i,1}(x)=x^{k_i-1}.
\end{equation}
Using Eqs. (2) and (3), we obtain,
\begin{equation}\label{eq30}
G_{i,0}(f,x)=(fx+1-x)^{k_i},
\end{equation}
\begin{equation}\label{eq31}
G_{i,1}(f,x)=(fx+1-x)^{k_i-1}.
\end{equation}

For a single network $i$ we obtain $g_i(x)=1-G_{i,0}(f_i,x)$, where
$f_i$ satisfies the equation $f_i=G_{i,1}(f_i,p)$. For loopless NON
of $n$ networks, we can obtain
\begin{equation}\label{eq32}
\mu'_{i}=p\prod\limits_{j=1,j \neq i}^{n}
[1-(f_j\mu'_j+1-\mu'_j)^{k_j}],
\end{equation}
where $f_j$ satisfies
\begin{equation}\label{eq33}
f_j=[f_j\mu'_j+1-\mu'_j]^{k_j-1}.
\end{equation}
Thus we can obtain
\begin{equation}\label{eq34}
P_{\infty} \equiv
\mu_{\infty}=p\prod\limits_{j=1}^{n}(1-f_j^{\frac{k_j}{k_j-1}}).
\end{equation}

where $f_i$ satisfies
\begin{equation}\label{eq35}
f_i=[(f_i-1)p\prod\limits_{j=1,j \neq i}^{n}
(1-f_j^{\frac{k_j}{k_j-1}})+1]^{k_i-1}.
\end{equation}

When all $n$ networks have the same degree $k$ i.e., $k_i=k$
($i=1,2,...,n$), we introduce a new variables $r=f^{\frac{1}{k-1}}$
into Eq. (\ref{eq35}), and the $n$ equations are reduced to a single
one
\begin{equation}\label{eq36}
r = (r^{k-1}-1)p(1-r^{k})^{n-1}+1,
\end{equation}
which can be solved graphically for any $p$. The critical case
corresponds to the tangential condition. Thus, we obtain that the
value of $r$ satisfies a transcendental equation
\begin{equation}\label{eq37}
1=(1-r)r^{k-2}[\frac{(n-1)kr}{1-r^{k}}+\frac{k-1}{1-r^{k-1}}].
\end{equation}
Solving $r$ from Eq. (\ref{eq37}), we can obtain the critical value
of $p_c$ and the the value of $P_{\infty}$ at $p_c$ as

\begin{equation}\label{eq38}
p_c = \frac{r-1}{(r^{k-1}-1)(1-r^{k})^{n-1}},
\end{equation}
The numerical solutions are shown in Fig. 7(b).

\begin{equation}\label{eq39}
P_{\infty} = p_c(1-r^{k})^n.
\end{equation}
We can obtain $P_{\infty}$ as a function of $r$ by substituting $r$
into Eq. (\ref{eq36}),
\begin{equation}\label{eq40}
 P_{\infty} = p \big \{1-\{p^{\frac{1}{n}}P_{\infty}^{\frac{n-1}{n}}
[(1-(\frac{P_{\infty}}{p})^{\frac{1}{n}} )^{\frac{k-1}{k}}-1 ]+1\}^k
\big \}^n.
\end{equation}

The results to compare the simulation and theory are shown in Fig.
6(a). The numerical simulation results is shown in Fig. 8(b) and
9(b).

For $n \gg 1$, Eq. (\ref{eq37}) can be rewritten as
\begin{equation}\label{eq41}
r=(\frac{1}{kn})^{\frac{1}{k-1}}.
\end{equation}
From Eq. (\ref{eq38}) and Eq. (\ref{eq41}) for the case when $n \gg
1$, we obtain \cite{bashan2011}
\begin{equation}\label{eq42}
p_c=\frac{1-r}{e^{r/k}},
\end{equation}
where $r$ satisfies Eq. (\ref{eq41}).

Since $k/(1-k)<0$ for $k>1$, it follows, in contrast to the ER case,
in the RR NON case $p_c$ can never be greater or equal to 1. This
shows that an ER NON is extremely more vulnerable compared to RR
NON, due to the critical role played in the ER by singly connected
nodes.

\section{ANALYTICAL RESULTS FOR THE CASE OF NON OF $n$ scale-free NETWORKS}
Here we study the case of a tree-like NON composed of $n$
scale-free(SF) networks. The generating function of each network is

\begin{equation}\label{eq43}
G_{i,0}(x) =
\frac{\sum^M_m[(k+1)^{1-\lambda}-k^{1-\lambda}]x^k}{(M+1)^{1-\lambda}-m^{1-\lambda}},
\end{equation}
\begin{equation}\label{eq44}
G_{i,1}(x) =
\frac{\sum^M_m[(k+1)^{1-\lambda}-k^{1-\lambda}]kx^{k-1}}{\sum^M_m[(k+1)^{1-\lambda}-k^{1-\lambda}]k},
\end{equation}
and
\begin{equation}\label{eq45}
G_{i,0}(f,x) = G_{i,0}(1+xf-x),
\end{equation}
\begin{equation}\label{eq46}
G_{i,1}(f,x) = G_{i,1}(1+xf-x).
\end{equation}

\begin{equation}\label{eq47}
g_i = 1-G_{i,0}(1+xf-x).
\end{equation}
\begin{equation}\label{eq48}
f_i = G_{i,1}(1+xf-x).
\end{equation}

Substituting Eq. (\ref{eq45}) - (\ref{eq48}) into Eqs.
(\ref{eq6})-(\ref{eq8}), we obtain

\begin{equation}\label{eq49}
x = p \prod_{j \neq i} \Big
\{1-\frac{\sum^{M_j}_{m_j}[(k+1)^{1-\lambda}-k^{1-\lambda}](1+f_jx_j-x_j)^k}{(M+1)^{1-\lambda}-m^{1-\lambda}}
\Big\},
\end{equation}

\begin{equation}\label{eq50}
f_j=\frac{\sum^{M_j}_{m_j}[(k+1)^{1-\lambda}-k^{1-\lambda}]k(1+f_jx_j-x_j)^{k-1}}{\sum^M_m[(k+1)^{1-\lambda}-k^{1-\lambda}]k},
\end{equation}
and
\begin{equation}\label{eq51}
P_{\infty} = p \prod^n_{j=1} \Big
\{1-\frac{\sum^{M_j}_{m_j}[(k+1)^{1-\lambda}-k^{1-\lambda}](1+f_jx_j-x_j)^k}{(M+1)^{1-\lambda}-m^{1-\lambda}}
\Big\}.
\end{equation}

Figs. 8(c) and 9(c) show the solutions for $P_{\infty}$ for several
values of $n$ and $m$ respectively.

\section{The case when $n\rightarrow \infty$} The behavior of the
networks for $n\to \infty$ depends only on $P(k)$, for small $k$.

If $P(1)+P(0)>0$, then no matter how large is $\langle k\rangle$ and
what is the degree distribution for the rest of $k$, the networks
completely collapse for large enough $n$ ($P_\infty=0$).

If $P(1)+P(0)=0$ the networks survive ($P_{\infty}>0$) for any $n$
and large enough $p<1$. Indeed, if $P(2)>0$ but $P(1)+P(0)=0$,
\begin{equation}\label{addeq3}
z_c \sim \frac{1}{n}(\frac{1}{2P(2)}-\frac{1}{\langle k\rangle}),
\end{equation}
which corresponds to the maximum of the r.h.s of Eq. (\ref{addeq1})
and $p_c\to 1$, $P_{\infty} \to 1$ can be found from Eqs.
(\ref{addeq1}) and (\ref{addeq2}). Thus, $p_c$ can not be greater
than 1, meaning that for all $n$ values, the NON is stable.

Assuming $\sum_0^{\ell-1}P(k)=0$ and $P(\ell)>0$ then,
\begin{equation}\label{addeq4}
z_c \sim [1/(n\ell P(\ell))]^{1/(\ell-1)}.
\end{equation}
and the way $p_c\to 1$ and $P_\infty \to 1$ can be found from Eqs.
(\ref{addeq1}) and (\ref{addeq2}) to be.

$P_\infty =1-C_1/n^{1/(\ell-1)}$ and $p_c=1-C_2/n^{1/(\ell-1)}$,
where $C_1>0$ and $C_2>0$ are constants, that can be easily found
from Eqs. (\ref{addeq1}) and (\ref{addeq2}).

For $P(2)>0$, $C_2=2P(2)/\langle k\rangle^2$ and
$C_1=(1-2P(2)/\langle k\rangle)^2/(2P(2))$  and for

$P(2)=0$, $C_2=(\ell-1)/(\ell P(\ell))^{1/(\ell-1)}$ and
$C_1=1/(\ell P(\ell))]^{1/(\ell-1)}$. Thus, when $P(0)+P(1)=0$, the
NON is stable for all $n$ ($p_c(n)<1$) and a condition for a minimal
$k(n)$, such as in Eq. (\ref{eq21}) does not exist.

\section{Conclusion}

In summary, we have developed a framework, Eqs.
(\ref{eq7})-(\ref{eq8}), for studying percolation of NON from which
we derived an exact analytical law, Eqs.~(\ref{eq22}) [for ER
networks] and (\ref{eq40}) [for RR networks], for percolation in the
case of a network of $n$ coupled networks. In particular for any $n
\geq 2$, cascades of failures naturally appear and the phase
transition becomes first order transition compared to a second order
transition in the classical percolation of a single network. These
findings show that the percolation theory of a single network is a
limiting case of a more general case of percolation of
interdependent networks. Due to cascading failures which increase
with $n$, vulnerability significantly increases with $n$. We also
find that for any tree-like network of networks the critical
percolation threshold and the mutual giant component depend only on
the number of networks and not on the topology (see Fig.~2). We
discuss the case for $n$ coupled ER networks, RR networks and SF
networks. We find that there exist the minimum $k$ to make the NON
survives, but no parameter for the RR and SF networks.

%\newpage
%\vspace*{-0.3cm}

%%%%%%%%%%%%%%%%%%%%%%%%%%%%% REFERENCES %%%%%%%%%%%%%%%%%%%%%%%%%

\section{Appendix}

\subsection{The case when $n\rightarrow \infty$ for ER NON}

Eq. (13) can be written as
\begin{equation}\label{aeq23}
n= \frac{f_c-1}{f_c \ln{f_c}}.
\end{equation}
Then we can get,
\begin{equation}\label{aeq24}
\frac{dn}{df_c}=\frac{\ln{f_c}+1-f_c}{(f_c\ln{f_c})^2}.
\end{equation}
When $n=1$,$f_c=1$, and $\frac{dn}{df_c}=0$; when $n>1$,$f_c<1$, and
$\frac{dn}{df_c}<0$.

So $f_c$ is a decreasing function of $n$ when $n>1$. We introduce a
new variable
\begin{equation}\label{aeq25}
\gamma=\frac{1-f_c}{nf_c},
\end{equation}
so, $f_c=e^{-\gamma}$. When $n\rightarrow\infty$, $f_c\rightarrow0$
and $\gamma\rightarrow\infty$. So studying the case
$n\rightarrow\infty$ is the same as studying the case
$\gamma\rightarrow\infty$.

Substitute Eq. (20) to Eq. (19), we obtain
\begin{equation}\label{aeq26}
n=\frac{1-e^{-\gamma}}{\gamma e^{-\gamma}}.
\end{equation}
We study $n$ as a function of $\gamma$ when
$\gamma\rightarrow\infty$,
\begin{equation}\label{aeq27}
\lim_{\gamma\rightarrow\infty}{n}=\frac{e^{\gamma}}{\gamma},
\end{equation}
\begin{equation}\label{aeq28}
\ln{n}=\gamma-\ln{\gamma}.
\end{equation}
Substituting Eq. (21) to Eq. (16) and consider the case when
$\gamma\rightarrow\infty$, we obtain
\begin{equation}\label{aeq29}
k_{\min}=\frac{\gamma}{(1-e^{-\gamma})^n},
\end{equation}
\begin{equation}\label{aeq30}
\lim_{\gamma\rightarrow\infty}{k_{\min}}=\frac{\gamma}{e^{\frac{1}{\gamma}}}.
\end{equation}
Substituting Eq. (23a) to Eq. (22b) and consider the case when
$\gamma\rightarrow\infty$, we obtain
\begin{equation}\label{aeq31}
k_{\min}=e^{-\frac{1}{\gamma}} \ln{n} + e^{-\frac{1}{\gamma}}
\ln{k_{\min}}+\frac{1}{\gamma}e^{-\frac{1}{\gamma}},
\end{equation}
\begin{equation}\label{aeq32}
\lim_{\gamma\rightarrow\infty}{k_{\min}}=\ln(n\ln{n})+\xi,
\end{equation}
where $\xi=O(\ln(\ln{n}))$. We can also obtain that when
$n\rightarrow\infty$, $p_c$ satisfies
\begin{equation}\label{aeq33}
p_c=\frac{1}{k}\ln(n\ln{n})+\xi,
\end{equation}
where $\xi=O(\ln(\ln{n}))$. This result is corresponds to large $n$
values in Fig. 7(a).

\end{document}